\documentclass[5p]{elsarticle}

\usepackage[english]{babel}
\usepackage[utf8x]{inputenc}
\usepackage[T1]{fontenc}
\usepackage{siunitx}

\DeclareSIUnit \parsec {pc}

\usepackage{cleveref}
\usepackage{amsmath}
\usepackage{graphicx}
\usepackage{subfig}
\usepackage[colorinlistoftodos]{todonotes}
\usepackage[colorlinks=true, allcolors=blue]{hyperref}
\usepackage{csvsimple,booktabs}
\usepackage{floatrow}

\addto\extrasenglish{%
}
\addto\extrasenglish{%
}

\addto\extrasenglish{%
}

\addto\extrasenglish{%
}

\definecolor{mygray}{gray}{0.6}

    \makeatletter
    \def\ps@pprintTitle{%
       \let\@oddhead\@empty
       \let\@evenhead\@empty
       \let\@oddfoot\@empty
       \let\@evenfoot\@oddfoot
    }
    \makeatother

\begin{document}
\begin{frontmatter}

\title{Sensitivity of the PICO-500 Bubble Chamber to Supernova Neutrinos Through Coherent Nuclear Elastic Scattering}

\author[UofA]{Tetiana Kozynets\corref{cor1}}
	\cortext[cor1]{kozynets@ualberta.ca}
\author[UofA]{Scott Fallows}
\author[UofA]{Carsten B. Krauss}
\address[UofA]{Department of Physics, University of Alberta, Edmonton, T6G 2E1, Canada}

\begin{abstract}
Ton-scale direct dark matter search experiments should be sensitive to neutrino-induced recoil events from either $^8$B solar neutrinos or the brief but intense flux from a core collapse supernova in the Milky Way. These low-threshold detectors are sensitive to the very low recoil energies, of order \SI{10}{\keV}, deposited via coherent elastic scatters between supernova neutrinos and target nuclei. Large superheated fluid detectors like PICO-500, a bubble chamber to be initially filled with an active target of \SI{1}{\tonne} of C$_3$F$_8$, should see multiple-bubble events from CE$\nu$NS if the detector is live during a neutrino burst from a supernova at a distance up to \SI{10}{\kilo\parsec}. This paper discusses conditions under which bubble chambers could be used as an independent measurement in the event of a supernova similar to SN 1987A, with particular sensitivity to the currently less-constrained heavy-lepton $\nu_x$ channel.
\end{abstract}

\end{frontmatter}

\section{Introduction}

At the end of their lifetime, massive stars, typically with $10M_{\odot} \leq M \leq 20M_{\odot}$, may undergo core collapse and explode as Type II supernovae, having most of the gravitational binding energy of their remnants radiated away in neutrinos \cite{bahcall1989neutrino}. Given a sufficiently small distance to a supernova progenitor and an adequate detection sensitivity, it is possible for these neutrinos to be witnessed on Earth, as they may engage in both charged current (CC) and neutral current (NC) interactions with matter. Any such observation is highly valuable in terms of contributing to the current understanding of core collapse and neutrino physics \cite{scholberg_supernova_2012}.  For that reason, since the detection of nearly a dozen neutrinos from SN 1987A in the Large Magellanic Cloud by the Kamiokande-II \cite{kamiokande_1987}, Irvine-Michigan-Brookhaven (IMB) \cite{bionta_IMB_1987}, and Baksan \cite{alekseev_baksan_1987} experiments, much effort has been put into planning and commissioning detectors capable of observing a large neutrino signal from the next supernova. Present facilities include the water Cherenkov detectors Super-Kamiokande \cite{ikeda_superkamiokande_2007} and IceCube \cite{abbasi_icecube_2011}, the scintillator-filled KamLAND \cite{kamland_collaboration_first_2003} and Borexino \cite{cadonati_supernova_2002}, as well as the Pb-based HALO \cite{duba_halo_2008}. The recently observed coherent elastic neutrino-nucleus scattering (CE$\nu$NS) of neutrinos of all flavors \cite{akimov_observation_2017,barranco_probing_2005}, which features large cross sections and constitutes the dominant part of the neutrino scattering rate, cannot be presently observed in the aforementioned experiments, as typical recoil energies of CE$\nu$NS interaction are of keV scale and therefore fall below the energy thresholds in these detectors.  Ton-scale direct dark matter search detectors, on the contrary, tend to have thresholds on the order of a few keV and may get around this limitation. In particular,  CLEAN \cite{horowitz_supernova_2003}, XMASS \cite{abe_detectability_2017}, LZ \cite{khaitan_supernova_2018}, and XENON1T \cite{lang_supernova_2016} were shown to have sufficient CE$\nu$NS sensitivity to detect neutrinos from supernovae within the Galaxy. The purpose of the present paper is to draw attention to the potential of a direct dark matter search bubble chamber to likewise detect supernova neutrinos via CE$\nu$NS using the example of the funded PICO-500 experiment, both for its planned initial configuration of a \SI{1}{\tonne} C$_3$F$_8$ target and for other similar configurations that could potentially be built and operated.

\subsection{Operational Principles of Superheated Fluid Detectors}

Bubble chambers use superheated fluids such as C$_3$F$_8$ to detect elastic scatters on target nuclei. Nuclear recoils that deposit an energy above the detector's thermodynamic threshold, as set by its operating temperature and pressure, will nucleate a bubble that grows to visible size. This visible bubble is detected by high-speed cameras that trigger a hydraulic compression of the target fluid until the pressure is high enough that the vapor condenses back into the liquid state. To reset the detector, the hydraulic system reduces the pressure, returning the target to the superheated state. The period during which the fluid is highly compressed after each bubble event defines a ``dead time'' when the detector is not sensitive to particle interactions. Superheated fluids have a very strong intrinsic rejection of electron recoils, typically of order 10$^{10}$ or larger, which can be adjusted as required by varying the thermodynamic conditions. The resulting very low overall event rates lead to typical live-time fractions of $\sim$90\%, as shown during the operation of the PICO-60 chamber filled with \SI{52}{kg} of C$_3$F$_8$ \cite{pico2l_prd,pico60_prl}.

The fluoroalkane fluids used in these chambers are typically inexpensive, and the detector technology is both conceptually and mechanically simple. This makes it quite practical to scale to large target masses, relative to other dark matter direct detection technologies. If the bubble chamber technique is combined with a high-neutron-density target, the flavor-blindness of CE$\nu$NS enables such chambers to have particularly good sensitivity, relative to large water Cherenkov neutrino detectors, to the heavy-lepton $\nu_{x}$ component of the supernova neutrino flux.

\subsection{PICO-500 Design and Neutrino Physics Reach}

PICO-500 is a ton-scale bubble chamber to be deployed underground at SNOLAB that is currently in the design and early procurement stage. This detector's greatly increased volume will push the boundary for current low-background bubble chamber technology. The baseline design is for a target composed of \SI{1}{\tonne} of superheated C$_3$F$_8$, but several other target options are being explored, both for this and for future chambers.
Stable operation of a bubble chamber filled with \SI{36.8}{\kg} of CF$_3$I was demonstrated during the first operational phase of PICO-60 \cite{pico60_prd1}. 
An even larger chamber filled with an iodine-rich target is of continued interest. Projections of such a chamber's sensitivity to supernova neutrinos via CE$\nu$NS are given in \autoref{sec:scattering_rates}, as compared to an equal volume of C$_3$F$_8$ and of the noble liquids $^{40}$Ar and $^{132}$Xe. We note that PICO-500 will not be sensitive to the CC channels accessible to classic neutrino detectors, as electronic interactions are invisible to the bubble chamber in the dark matter operational mode. 

For the future PICO-500 operation, the goal is to restrict the background from neutron induced multi-bubble events to a few per year. The appearance of a multiple-bubble event, such as that expected from nearby-supernova CE$\nu$NS, is easily distinguishable both from PICO's single-bubble WIMP-search channel and from its simultaneous multi-bubble neutron channel. This detection would be a highly complementary signal to those seen in coincidence in other detectors.

\section{Inputs to Sensitivity Projection}
\subsection{Supernova Neutrino Fluxes}\label{sec:sne_fluxes}

A three-temperature Boltzmann distribution model has been widely employed to describe the simulated spectra of supernova neutrinos \cite{horowitz_supernova_2003, amaya_study_2012}: 

\begin{equation}
\label{eq:boltzmann_spectra}
\psi^{(\mathrm{B})}_j(E_\nu) = \frac{1}{4\pi d^2} \cdot \frac{n_j E_{\nu}^2}{2\tau_j^3} \cdot e^{-\frac{E_{\nu}}{\tau_j}},
\end{equation}
where $n_j$ denotes the number of neutrinos of flavor $j$ ($\nu_e,~ \bar{\nu}_e$, or $\nu_x \equiv \{ \nu_{\tau}, \bar{\nu}_{\tau}, \nu_{\mu}, \bar{\nu}_{\mu} \}$) emitted from a supernova at a distance $d$ from the Earth, $\tau_j$ stands for the effective temperature specific for this flavor, and $E_{\nu}$ is the neutrino energy. The result of the right-hand side expression evaluation is then the neutrino number flux density $\psi^{(B)}_j(E_\nu)$, which has the units of inverse energy if temperature is taken in energy units.

In the course of the past decade, more accurate models have been developed. The current standard is the pinched flux model, involving the ``pinching parameter'' $\alpha$ and the average neutrino energy $\left\langle E_{\nu}\right\rangle$ \cite{keil_monte_2003,tamborra_high-resolution_2012}:
\begin{equation}
\label{eq:scholberg_spectra}
\psi^{(\mathrm{p})}(E_{\nu}, \mathcal{E}) = \frac{\mathcal{E}}{4 \pi d^2 \left\langle E_{\nu} \right\rangle} \cdot f^{\mathrm{(p)}}(E_{\nu}),
\end{equation}
where
\begin{equation}
f^{\mathrm{(p)}}(E_{\nu}) = A  \left(\frac{E_{\nu}}{\left\langle E_{\nu}\right\rangle} \right)^{\alpha} \exp \left[-(\alpha + 1) \frac{E_{\nu}}{\left\langle E_{\nu}\right\rangle}\right]
\end{equation}
is the normalized gamma distribution ($A = \frac{(\alpha + 1)^{\alpha + 1}}{\left\langle E_{\nu}\right\rangle \Gamma(\alpha + 1)}$) and $\mathcal{E}$ is the total energy emitted in neutrinos \cite{migenda_detecting_2016}. Both $\alpha$ and $\left\langle E_{\nu}\right\rangle$ are in general time-dependent, and their values can be obtained from simulations or fits to future high-resolution spectral data. With $\alpha = 2$ at all post-bounce times, \autoref{eq:scholberg_spectra} reduces to \autoref{eq:boltzmann_spectra} \cite{huedepohl_neutrino_2010}. 

In this work, we will consider only the pinched flux model and examine the time evolution of the corresponding spectrum using the simulated spherically symmetric (1D) neutrino signal from the Garching Core-Collapse Supernova Archive \cite{garching_archive}, obtained for a 20$M_{\odot}$ progenitor \cite{woosley_nucleosynthesis_2007} leaving a neutron star of nearly $1.95 M_{\odot}$ baryonic mass. This model is preferred over three-temperature Boltzmann because with $\alpha$ held constant, as implied by \autoref{eq:boltzmann_spectra}, the latter becomes an oversimplification when energy moments evolve in time. The dependence of $\left\langle E_{\nu} \right\rangle$ and $\left\langle E^2_{\nu} \right\rangle$ outputs of a 1D supernova simulation \cite{mirizzi_supernova_2015} on the post-bounce time $t_{\mathrm{pb}}$ enables us to find $\alpha (t_{\mathrm{pb}})$ directly from the ratio

\begin{equation}
\label{eq:alpha}
\frac{\left\langle E^2_{\nu} \right\rangle}{\left\langle E_{\nu} \right\rangle ^2} = \frac{2 + \alpha}{1 + \alpha}.
\end{equation}

\begin{figure}[h!]
  \includegraphics[scale=0.45]{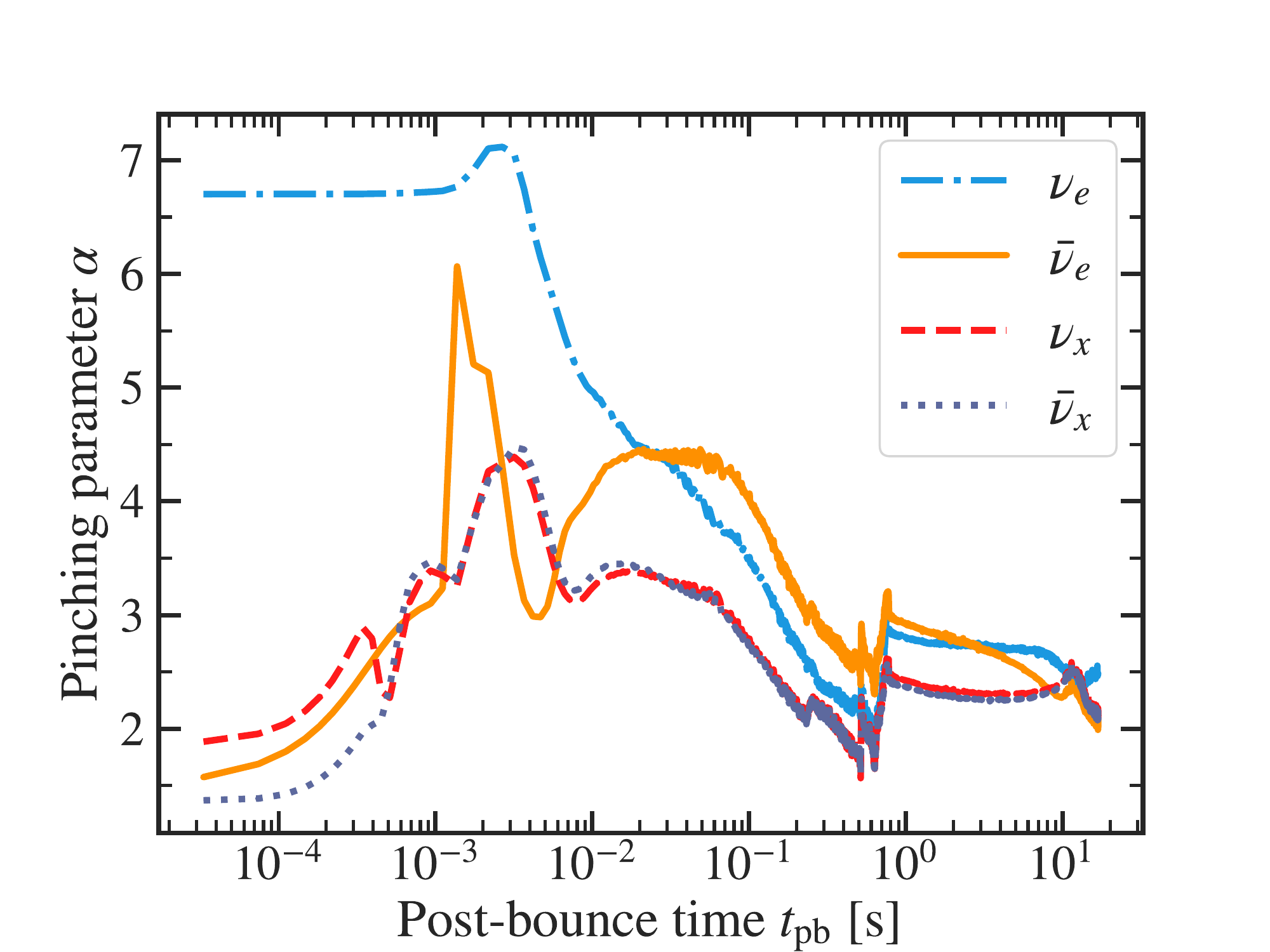}
  \centering
  \caption{Dependence of the parameter $\alpha$ (see \autoref{eq:scholberg_spectra}) on the post-bounce time $t_{\mathrm{pb}}$: Simulated neutrino signal \cite{garching_archive} from the explosion of a $20 M_{\odot}$ progenitor  \cite{woosley_nucleosynthesis_2007,mirizzi_supernova_2015}.}
  \label{fig:alpha}
\end{figure}

\begin{figure}[h!]
  \includegraphics[scale=0.45]{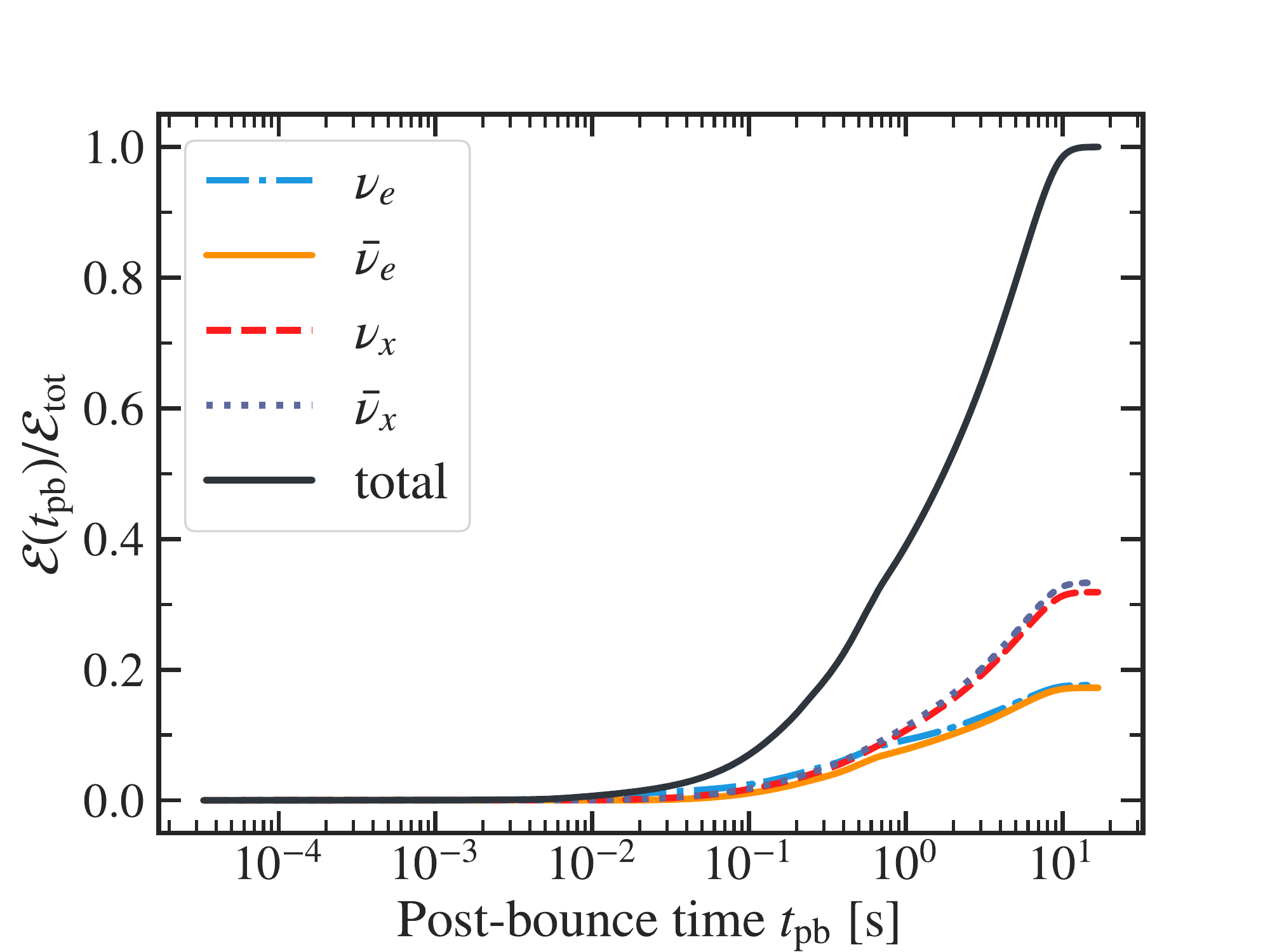}
  \centering
  \caption{Fraction of the total energy $\mathcal{E}_{\mathrm{tot}}$ radiated in time $t_{\mathrm{pb}}$ as a function of $t_{\mathrm{pb}}$: Simulated neutrino signal \cite{garching_archive} from the explosion of a $20 M_{\odot}$ progenitor \cite{woosley_nucleosynthesis_2007,mirizzi_supernova_2015}.}
  \label{fig:enfractions}
\end{figure}

In \autoref{fig:alpha}, we reproduce the time dependence of the $\alpha$ parameter for different flavors (with $\nu_x \equiv \{\nu_\mu, \nu_\tau\}$ and $\bar{\nu}_x \equiv \{\bar{\nu}_\mu, \bar{\nu}_\tau\}$, as defined in \cite{mirizzi_supernova_2015}). 
We observe that before $t_{\mathrm{pb}}$ reaches \SI{1}{s}, $\alpha$ changes drastically with time and stays far away from the Boltzmann-like $\alpha = 2$ for all flavors. Higher values of this parameter for $\nu_e$ and $\bar{\nu}_e$ imply more severe pinching than for $\nu_x$ and $\bar{\nu}_x$. Nearly half of the total energy is, however, radiated within the first two seconds following the core collapse. To show this, we plot the ratio of the luminosity integrated up to $t_{\mathrm{pb}}$ to the total energy of nearly $4.3 \times 10^{53}$ \SI{}{ergs} produced over the time scale of \SI{16.8}{s} as a function of $t_{\mathrm{pb}}$ in \autoref{fig:enfractions}. We note that since the explosion of the star was triggered artificially in the discussed 1D model \cite{mirizzi_supernova_2015}, the neutrino emission properties in the accretion phase (\SIrange[range-phrase=--,range-units=single]{0.2}{1}{s}) might differ from those of the full supernova models without any assumed symmetry (3D), which do explode naturally but are not currently run over time scales longer than \SI{0.5}{s} due to high computational costs. Once sufficiently long neutrino signals from 3D supernova simulations become available, neutrino emission in the accretion phase can be compared between the 1D and the 3D cases, and the results presented in the following sections can be further refined. However, as the exact time evolution of neutrino spectra is expected to differ even more across progenitor models and neutron star remnant masses than between 1D and 3D simulations, the present discussion is sufficient for a basic sensitivity projection study.

\begin{figure}[h!]
  \includegraphics[scale=0.45]{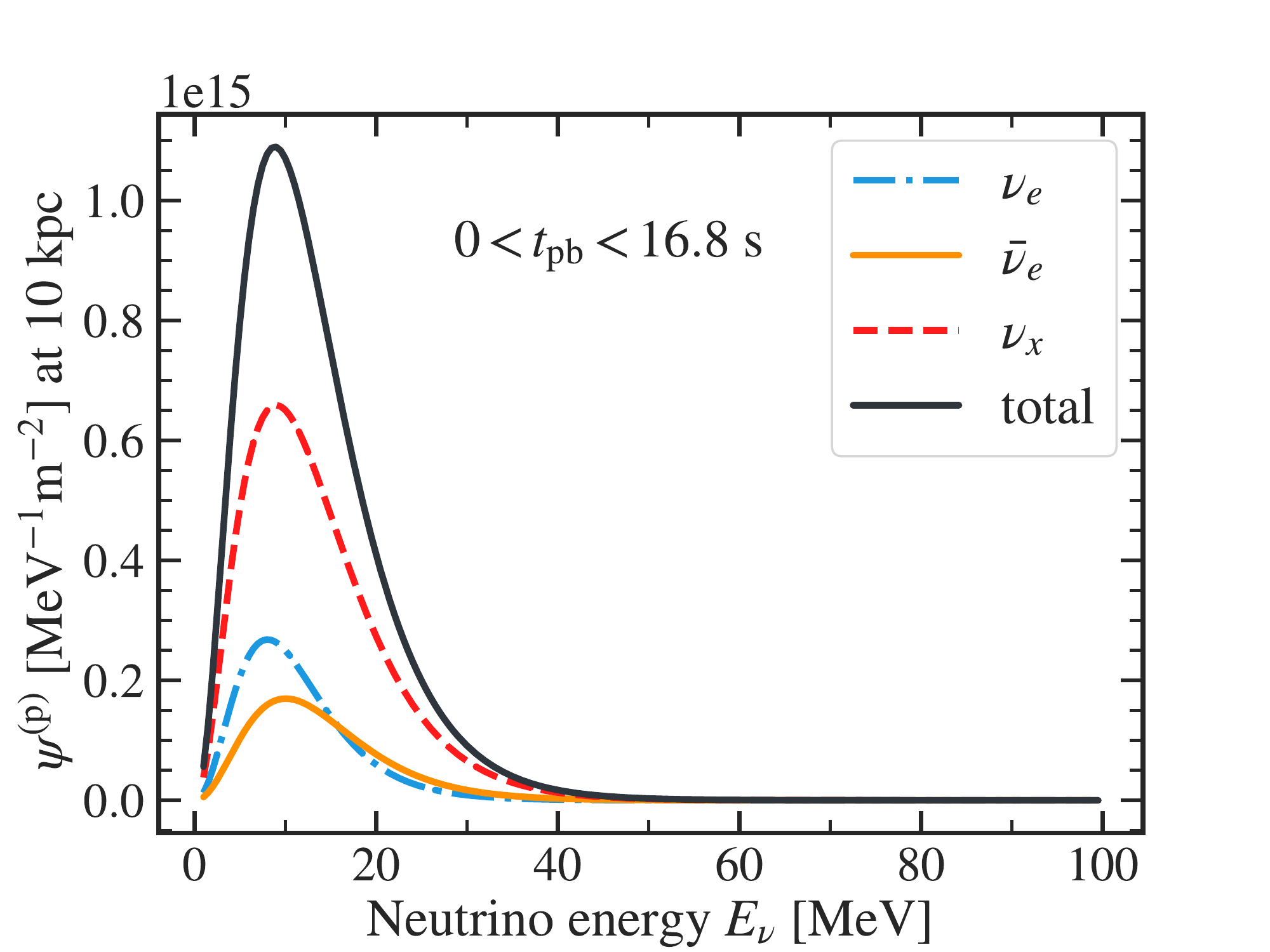}
  \centering
  \caption{Pinched neutrino spectrum (\autoref{eq:scholberg_spectra}) integrated from the bounce time up to \SI{16.8}{s} for a 20$M_{\odot}$ progenitor \cite{garching_archive,woosley_nucleosynthesis_2007,mirizzi_supernova_2015}.}
  \label{fig:pinched_and_boltzmann_spectra}
\end{figure}

The time-integrated pinched fluxes for different flavors are shown in \autoref{fig:pinched_and_boltzmann_spectra}, where $\nu_x$ now represents the sum of all heavy-lepton neutrino contributions. The integral runs up to \SI{16.8}{s} (full duration of the neutrino signal from \cite{garching_archive}). 

Predictions of the cumulative number of bubbles expected from a full post-bounce signal impose several clear constraints on the practical design of the experiment. The time during which the detector is to be held in the superheated state after the first detected bubble is a significant factor in planning its operational cycle. The goal is to capture as many scatter events as possible and yet recompress quickly enough to keep the detector's live fraction high. Therefore we are interested in the total scatter number as a function of post-bounce time, which we will denote as $t$ for the remainder of this section for brevity. To evaluate such time-dependent yields, we make use of the discrete set of $L_i (t_i)$ values from the Garching group simulation of the neutrino signal, where $L_i$ is the neutrino luminosity in the units of ergs per second, and employ the trapezoidal rule to integrate the luminosity within each $[t_i, t_{i+1}]$ bin:
\begin{equation}
\Delta\mathcal{E}(t_i) \equiv \int_{t_i}^{t_{i+1}}L(t) \mathrm{d}t \approx \frac{(L_i + L_{i+1}) \times (t_{i+1} - t_i)}{2},
\end{equation}
where we define $\Delta\mathcal{E}(t_i)$ to be the energy emitted in neutrinos in $\Delta t_i$ between $t_i$ and $t_{i+1}$ for a certain neutrino flavor. The total energy normalizes the pinched flux within this bin:
\begin{equation}
\label{eq:flux_normalization}
\Delta \psi^{\mathrm{(p)}}_i = \frac{\Delta \mathcal{E} (t_i) }{4 \pi d^2 \left\langle E_{\nu} \right\rangle} \cdot f^{\mathrm{(p)}}(E_{\nu}).
\end{equation} 
By numerically integrating the resulting flux rates, $\frac{\Delta \psi^{\mathrm{(p)}}_i}{\Delta t_i}$, over time, we obtain the full time dependence of the neutrino spectrum. To finally calculate the expected event rate for both time-integrated and time-dependent scenarios, it now only remains to introduce the $E_{\nu}$-dependence of the CE$\nu$NS cross sections (see \autoref{sec:cross_sections}). \autoref{sec:scattering_rates} and \autoref{sec:timedep_scattering_rates} present these estimates for PICO-500 filled with C$_3$F$_8$ as the proposed target liquid, as well as other liquids that could potentially fill the same chamber volume in the future.

\subsection{Elastic Scattering Cross Sections}\label{sec:cross_sections}

Following \cite{papoulias_coherent_2017}, we define the differential cross section of coherent elastic scattering as a function of neutrino energy $E_{\nu}$ and recoil energy $T$:
\begin{align}
\label{eq:diff_cross_section}
\frac{\mathrm{d}\sigma}{\mathrm{d}T} & = \frac{G_F^2}{4\pi} M \cdot \left[ (Q_W^V)^2  \left(1 - \frac{MT}{2E_{\nu}^2} \right) \right. \nonumber \\ 
& \left. + (Q_W^A)^2 \left(1 + \frac{MT}{2E_{\nu}^2} \right) \right] \cdot F(Q^2)^2.
\end{align}
In \autoref{eq:diff_cross_section}, $G_F \approx \SI{1.166}{\GeV}^{-2}$ is the Fermi coupling constant; $M = AM_N \approx (N+Z)\cdot \SI{931.5}{\MeV}$ is the mass of the target nucleus, with $N$ representing the number of neutrons, and $Z$ -- that of protons; and $F(Q^2)$ is the Helm-type ground-state elastic form factor \cite{papoulias_coherent_2017}

\begin{equation}
\label{eq:form_factor_eq}
F(Q^2) = \frac{3j_1(Q R_0)}{Q R_0} \exp\left[-\frac{1}{2} (Qs)^2\right],
\end{equation}
where momentum transfer $Q$ is related to the recoil energy $T$ via $Q = \sqrt{2MT}$ and $j_1$ is the first-order spherical Bessel function. The constant $R_0$ in \autoref{eq:form_factor_eq} is uniquely defined for a target nucleus with a given atomic mass $A$ by $R_0 = \sqrt{R^2 - 5s^2}$, where $R = 1.2 A^{1/3}$~fm is the effective nuclear radius and $s \approx \SI{0.5}{fm}$ is the nuclear skin thickness.

The other expressions appearing in \autoref{eq:diff_cross_section} are the vector and axial-vector nuclear charges, $Q_W^V$ and $Q_W^A$, such that
\begin{equation}
Q_W^V = N - (1 - 4 \sin^2\theta_W)\cdot Z,
\end{equation}
with the weak mixing angle $\sin^2\theta_W \approx 0.2386$ at momentum transfer $Q \leq$ \SI{100}{MeV} \cite{erler_weak_2005,doi:10.1146/annurev-nucl-102212-170556}, and 
\begin{equation}
\label{eq:axial_vector_charge}
Q_W^A = g_p^A(Z_{+} - Z_{-}) + g_n^A(N_{+} - N_{-}),
\end{equation}
where $g_p^A/g_n^A$ are the effective axial-vector coupling constants for neutral current neutrino-proton/neutron interactions, and $Z_{\pm}/N_{\pm}$ are the numbers of spin up ($+$) and down ($-$) protons/neutrons. In spin-zero nuclei, such as $^{12}$C, $^{40}$Ar and $^{132}$Xe, both terms in \autoref{eq:axial_vector_charge} are equal to 0, and we end up with $Q_W^A(\mathrm{^{12}C, ^{40}Ar, ^{132}Xe}) = 0$. For nuclei that carry a non-zero spin, such as $^{19}$F and $^{127}$I, it is typically the case that $\frac{Q_W^A}{Q_W^V} \sim \frac{1}{A}$ \cite{papoulias_coherent_2017}, which lets us safely neglect the axial-vector term for $^{127}$I but not for $^{19}$F. Therefore, we write for fluorine nucleus:
\begin{equation}
\label{eq:axial_vector_charge_Fluorine}
Q_W^A(^{19}\mathrm{F}) = g_p^A \approx 1.09,
\end{equation}
where we made use of the medium-suppressed neutrino-proton axial-vector coupling constant \cite{raffelt_self-consistent_1995}. 

\begin{figure}[h!]
    \subfloat[$\sigma(E_{\nu})$ for $\nu$ scattering off C and F in C$_3$F$_8$, with $T_{\mathrm{min}} = $ \SI{2}{\keV}]{\label{crossSec_sub1}\includegraphics[scale = 0.35]{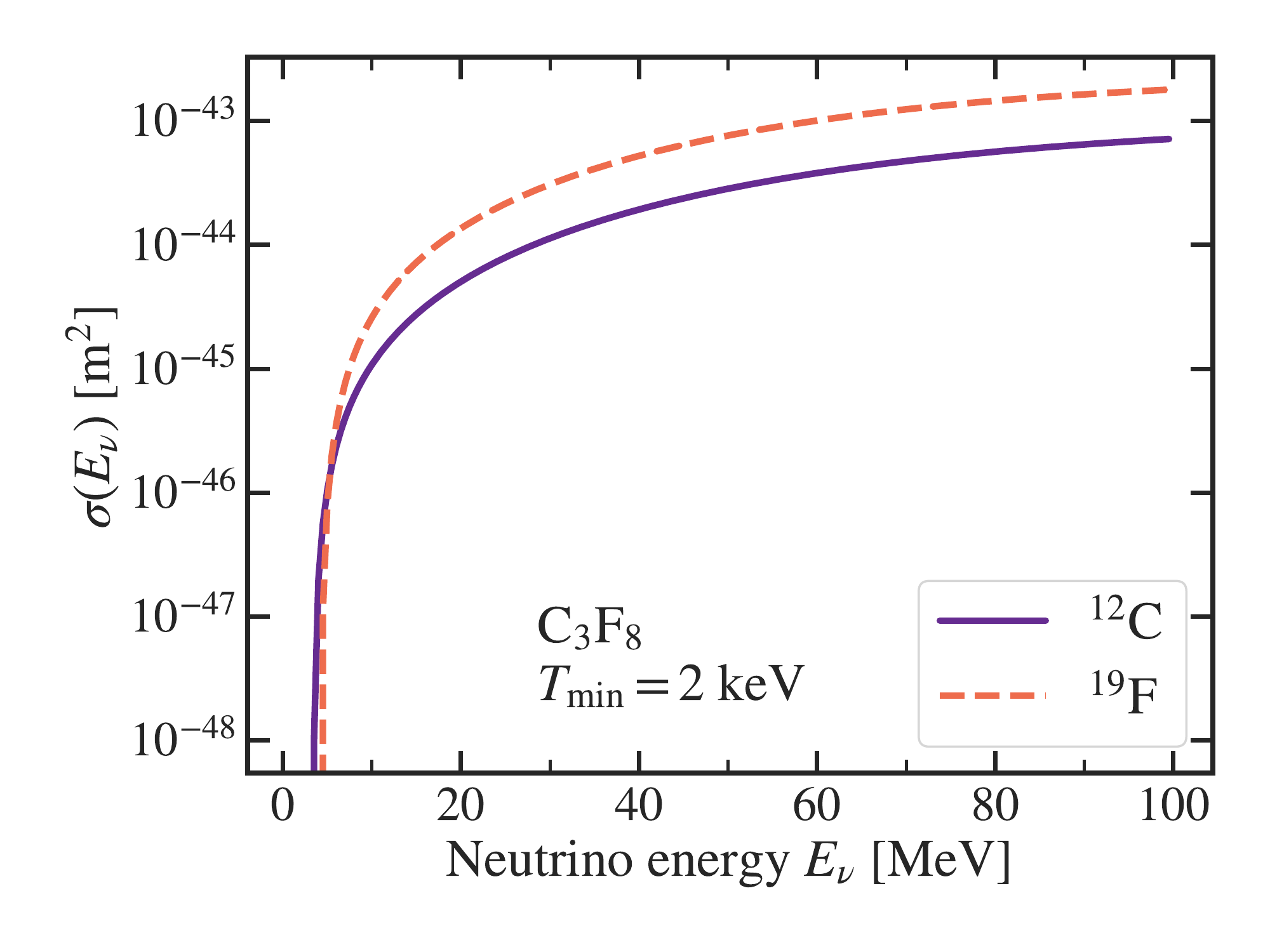}} \\
    \subfloat[$\sigma(E_{\nu})$ for $\nu$ scattering off C, F, and I in CF$_3$I, with $T_{\mathrm{min}} = $ \SI{10}{\keV}]{\label{crossSec_sub2}\includegraphics[scale = 0.35]{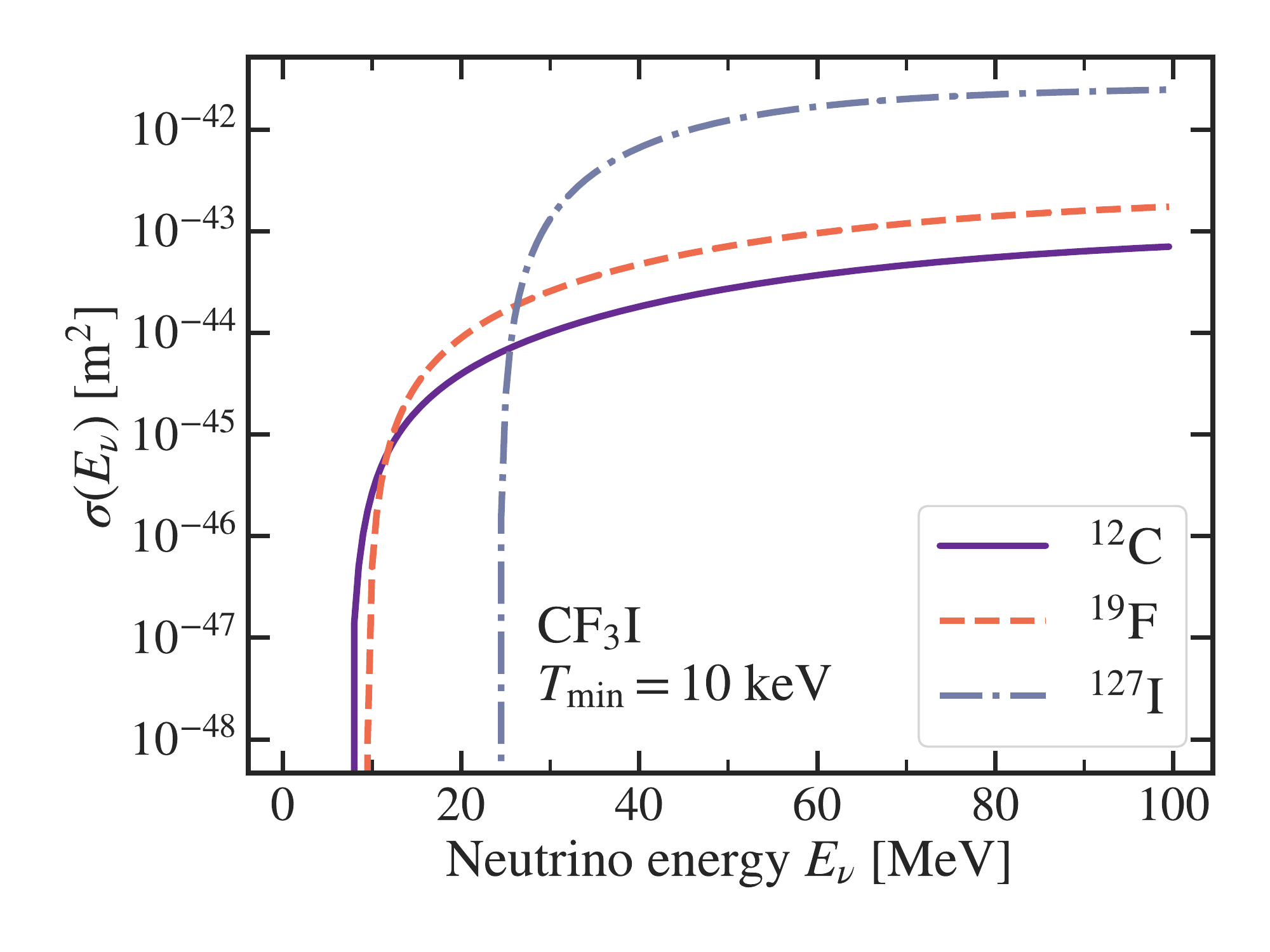}} \\ 
     \subfloat[$\sigma(E_{\nu})$ for $\nu$ scattering off $^{40}$Ar and $^{132}$Xe, with $T_{\mathrm{min}} = $ \SI{10}{\keV}]{\label{crossSec_sub3}\includegraphics[scale = 0.35]{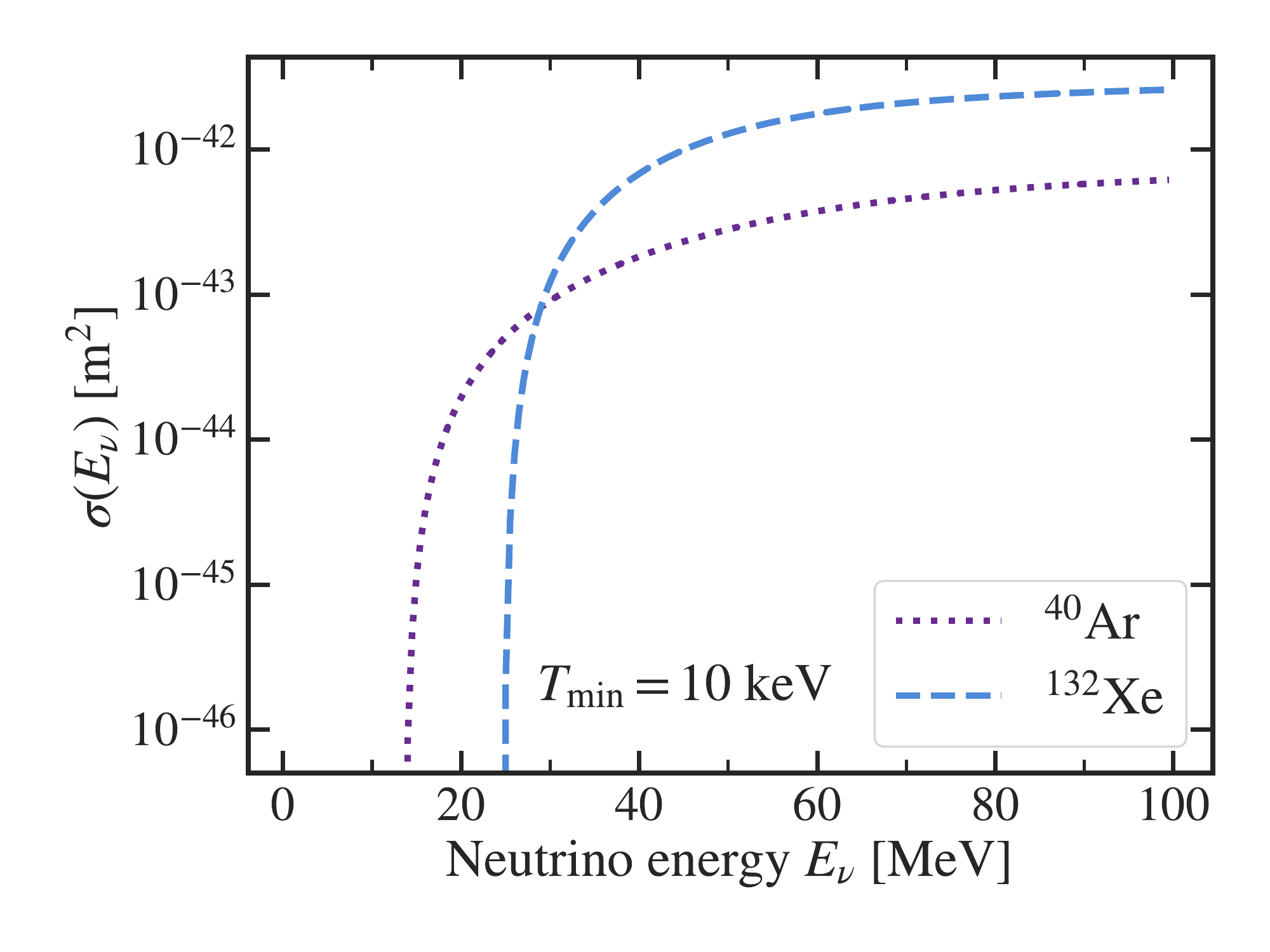}}
    \caption{Recoil energy integrated cross sections of coherent elastic neutrino scattering from C, F, I, Ar, and Xe nuclei plotted against neutrino energy $E_{\nu}$.}
    \label{fig:total_crossSec_full}
\end{figure}

For each $E_\nu$ in a discrete set $\{E_{\nu,0}, E_{\nu,1}, ..., E_{\nu,n} \}$ with $E_{\nu,0} \equiv E_{\mathrm{min}} =$ \SI{1}{MeV}, $E_{\nu,n} \equiv E_{\mathrm{max}} =$ \SI{100}{MeV}, and a constant step of \SI{0.5}{MeV}, we integrated the right-hand side in \autoref{eq:diff_cross_section} numerically with respect to $T$ from $T_{\mathrm{min}}$, standing for the detection threshold, to $T_{\mathrm{max}} \equiv \frac{2E_{\nu}^2}{M + 2E_{\nu}}$, representing the maximum recoil energy at a given neutrino energy $E_{\nu}$, to get full cross section $\sigma (E_{\nu})$ \cite{strigari_neutrino_2009}. 
\autoref{crossSec_sub1} shows these recoil energy integrated cross sections plotted against $E_{\nu}$ for C and F nuclei in C$_3$F$_8$ assuming a threshold of $\SI{2}{\keV}$, while \autoref{crossSec_sub2} presents the same for C, F, and I in CF$_3$I and Ar at a $\SI{10}{\keV}$ threshold.

\section{Results}
\subsection{Time-Integrated Scattering Rates}\label{sec:scattering_rates}

From the recoil energy integrated cross section $\sigma (E_{\nu})$ and the functional form of the neutrino flux density, we can evaluate the total number of scatters off $N$ target nuclei in the whole range of possible recoil energies as

\begin{equation}
\label{eq:total_yield}
N_{\mathrm{CE}\nu\mathrm{NS}} = N \times \int_{E_{\mathrm{min}}}^{E_{\mathrm{max}}} \sigma(E_{\nu}) \psi^{\mathrm{(p)}}(E_{\nu})\mathrm{d}E_{\nu}.
\end{equation}

\begin{figure}[h!]
    \includegraphics[scale = 0.45]{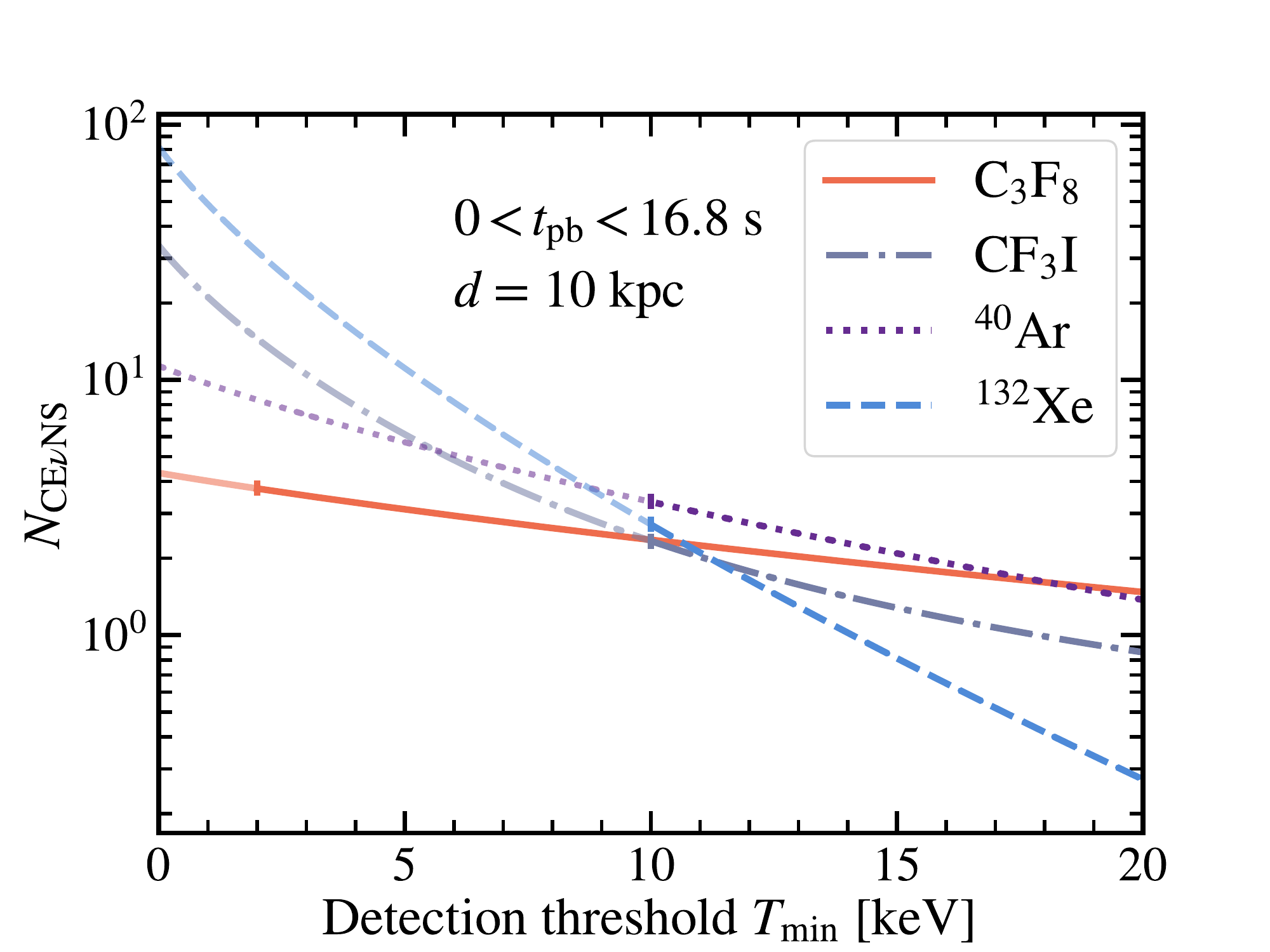}
    \centering
    \caption{Total number of neutrino-nucleus scatters in the volume of PICO-500 filled with different target liquids, evaluated using the pinched flux model (\autoref{eq:scholberg_spectra}) applied to the time-resolved 1D neutrino signal from a simulated explosion of a $20M_{\odot}$ progenitor \cite{mirizzi_supernova_2015}. The semi-transparent lines represent thresholds currently deemed inaccessible, and the vertical markers denote the upper boundaries of these regions. }
    \label{fig:pico500_threshold_dep}
\end{figure}

\autoref{tab:yield_table} lists the results of applying \autoref{eq:total_yield} to the case of Helm-type form factor (\autoref{eq:form_factor_eq}) and the pinched flux density model (\autoref{eq:scholberg_spectra}) for neutrinos in the range \SIrange[range-phrase=--,range-units=single]{1}{100}{\MeV} emitted from a supernova at \SI{10}{\kilo\parsec}, with different target liquids and detection thresholds considered for PICO-500. We show how the total number of scatters with recoil energies above the detection threshold $T_{\mathrm{min}}$ varies with the value of $T_{\mathrm{min}}$ in \autoref{fig:pico500_threshold_dep} for the total neutrino flux integrated until $t_{\mathrm{pb}}$ = \SI{16.8}{s}. 

These time-integrated values give insight into the sensitivity of the detector that could be achieved if its dead time was negligible during the supernova burst. For the reasons given in \autoref{sec:sne_fluxes}, a time-dependent treatment of the total yield is preferable to optimize the detector operation; this discussion follows in \autoref{sec:timedep_scattering_rates}. 

\begin{table*}[htb!]
\centering
\caption{Simulated total numbers of scatters that could be observed in \SI{725}{L} of superheated C$_3$F$_8$, CF$_3$I, $^{40}$Ar, and $^{132}$Xe liquids at different detection thresholds for a supernova at $d = $ \SI{10}{\kilo\parsec}, integrated over 16.8 seconds assuming the pinched flux model (\autoref{eq:scholberg_spectra}) for 1D neutrino signal data \cite{garching_archive}. For C$_3$F$_8$, the same results apply to the case of a \SI{25}{t} chamber and a \SI{50}{kpc} distance to a supernova. The values for thresholds currently deemed inaccessible are shown in gray.}
\label{tab:yield_table}
\begin{tabular}{lllllll}
\hline
Target  & $T > 0.5$ keV & $T > 2$ keV & $T > 5$ keV   & $T > 10$ keV & $T > 15$ keV & $T > 20$ keV \\ \hline
C$_3$F$_8$ [\SI{1}{t}] & \textcolor{mygray}{4.2}  & 3.8 & 3.1     & 2.4          & 1.8          & 1.5      \\
CF$_3$I [\SI{1.4}{t}]  & \textcolor{mygray}{26.2} & \textcolor{mygray}{14.5} & \textcolor{mygray}{6.1} & 2.3   & 1.3    & 0.9    \\
LAr [\SI{1.1}{t}]      & \textcolor{mygray}{10.5} & \textcolor{mygray}{8.4}  & \textcolor{mygray}{5.7}  & 3.3      & 2.1       & 1.4          \\
LXe [\SI{2.2}{t}]      & \textcolor{mygray}{62.1} & \textcolor{mygray}{31.9} & \textcolor{mygray}{11.1} & 2.7  & 0.8   & 0.3  \\ \hline
\end{tabular}
\end{table*}

\subsection{Time-Dependent Scattering Rates}\label{sec:timedep_scattering_rates}

\begin{figure}[h!]
  \includegraphics[scale=0.45]{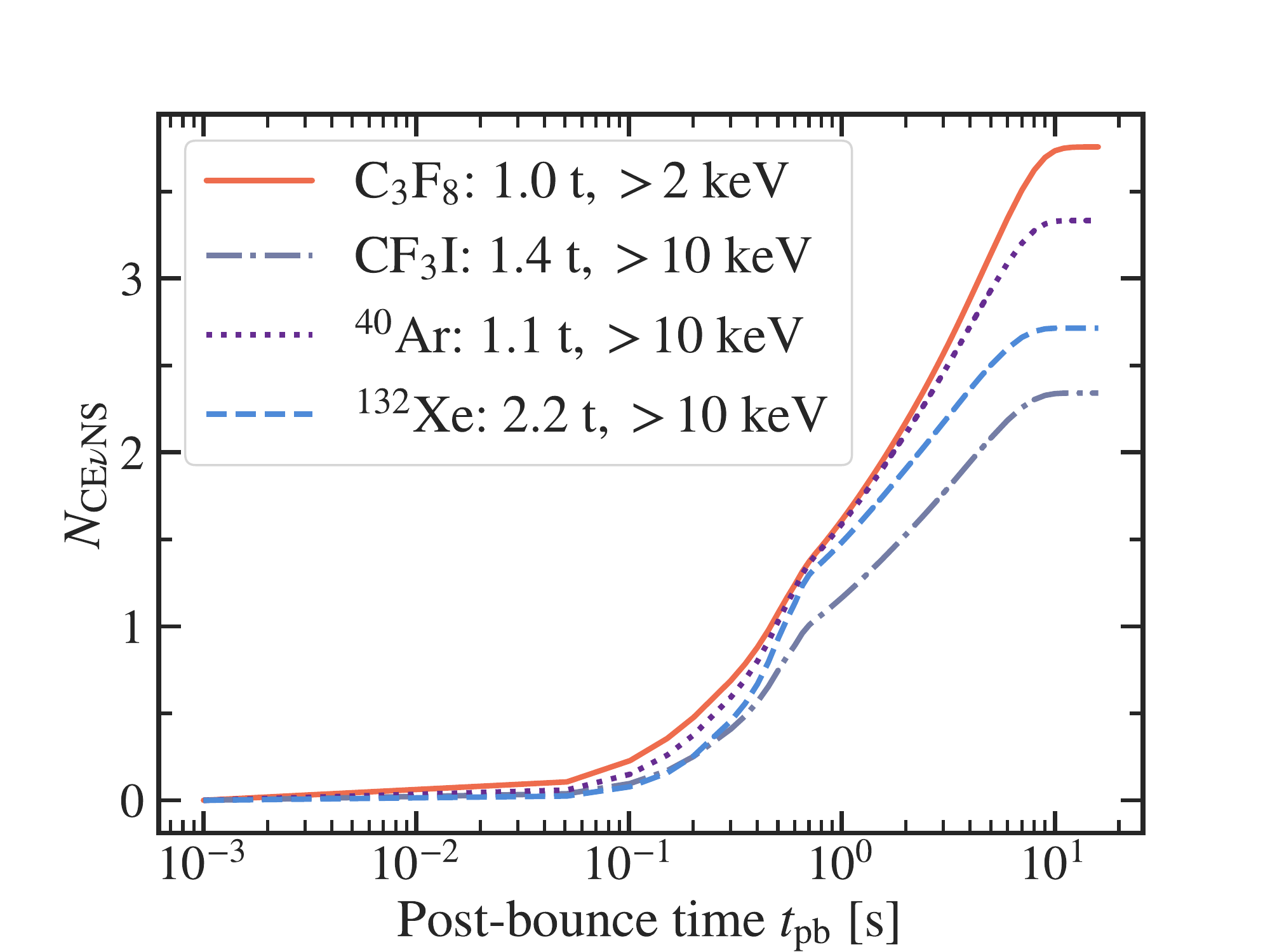}
  \centering
  \caption{Time dependence of the integrated event rates in C$_3$F$_8$ as the proposed PICO-500 target liquid and its possible future alternatives filling the same volume. The numerical values of yields integrated up to \SI{16.8}{s} are given in \autoref{tab:yield_table}.}
  \label{fig:time_dependence}
\end{figure}

Following the procedure described in \autoref{sec:sne_fluxes}, we evaluated the number of observable neutrino-nucleus scatters integrated until post-bounce time $t_{\mathrm{pb}}$ as a function of $t_{\mathrm{pb}}$. This time dependence is presented in \autoref{fig:time_dependence} for the four considered liquids and the respective minimum detection thresholds that can be practically achieved with each liquid at the time of writing.

From \autoref{fig:time_dependence}, we readily see that \SI{1}{t} of C$_3$F$_8$ with \SI{2}{keV} detection threshold turns out to give the highest yield among the investigated bubble chamber options at all post-bounce times. To quantify this statement, we provide the estimates of the time required to accumulate 1, 2, and 3 bubbles for each of C$_3$F$_8$, CF$_3$I, $^{40}$Ar, and $^{132}$Xe in \autoref{tab:time_toget_k_table}.

\begin{table*}[htb!]
\centering
\caption{Time since the supernova burst which is required to expect $k$ bubbles in C$_3$F$_8$ and other potential liquids for PICO-500 detector, with a minimum achievable detection threshold assumed for each liquid (see \autoref{tab:yield_table}).}
\label{tab:time_toget_k_table}
\begin{tabular}{lllll}
\hline
$k$  & $t$(C$_3$F$_8$, $k$) [s] & $t$(CF$_3$I, $k$) [s] & $t$($^{40}$Ar, $k$) [s]  & $t$($^{132}$Xe, $k$) [s] \\ \hline
1 & 0.5  & 0.7 & 0.5   & 0.6    \\
2  & 1.7 & 4.4 & 1.8 & 2.4 \\
3   & 4.5 & N/A & 6.0 & N/A  \\ \hline
\end{tabular}
\end{table*}

In this paper we assume that it is feasible for a bubble chamber to remain superheated for several seconds after the first nucleation. Preliminary results from the operation of a water-free test chamber \cite{rsu_drexel} with a similar design to PICO-500 indicate that significantly slower recompression is possible in this chamber configuration. However, the maximal duration of this compression delay needs to be experimentally determined in a large-scale bubble chamber. The goal of establishing sensitivity to processes on the order of seconds will impact the operational design for upcoming PICO detectors. 
 
\section{Discussion}\label{sec:discussion}

As per \autoref{tab:yield_table}, our expectation for the number of scatters observable from CE$\nu$NS of supernova neutrinos in PICO-500, whose current design includes a $\sim$\SI{725}{L} volume of C$_3$F$_8$, implies that such a detection is possible even at large thresholds. Indeed, for $T_{\mathrm{min}}$ = \SI{10}{\keV}, we would be able to confirm 2 CE$\nu$NS events if a supernova happened at a distance \SI{10}{\kilo\parsec} from the Earth and the detector was kept live for the whole neutrino signal duration. The latter scenario is far from realistic given the operational principles of bubble chambers, and yet the fact that thresholds as low as \SI{2}{\keV} can be reached for C$_3$F$_8$ with the current technology makes further investigation in this direction worthwhile. According to \autoref{tab:time_toget_k_table}, if the C$_3$F$_8$ remains superheated for 4 seconds after the first bubble is observed, all of the 3 expected scatters might be detected before recompression. Moreover, the second bubble is expected to be observed after 1.2 seconds after the detection of the first one. Time-separated multiple-bubble events would be highly distinctive from both the single-bubble events expected from the interaction of WIMPs and from the nearly simultaneous nucleations of multiple bubbles induced by neutron background. If supported by coincidence data from neutrino detectors, such a signal would thus act as solid confirmation of the neutrino emission from the supernova. While the PICO-500 data will not be sufficient to make inferences about the shapes of neutrino spectra on its own, the detection of even just a few events would add into the bigger picture of time distributions of neutrino fluxes collected at other sites. 

The results obtained for CF$_3$I, LAr, and LXe within the same volume are less promising in terms of detecting neutrinos from the same progenitor in a PICO-500-sized detector: indeed, we expect to detect only about 2 events above \SI{10}{\keV} in \SI{1.4}{\tonne} of CF$_3$I and 3 events in \SI{1.1}{\tonne} of LAr and \SI{2.2}{\tonne} of LXe above the same threshold, with the event rates integrated over \SI{16.8}{s}. The second CE$\nu$NS-caused bubble might, however, be seen in LAr just after \SI{1.3}{s} after the first one, which makes it the next most suitable liquid for the purpose of supernova neutrino detection in a chamber of this kind. The PICO Collaboration is actively considering the possibility of operating liquid argon chambers; in particular, Fermilab has recently awarded funding for a scintillating argon bubble chamber for CE$\nu$NS detection. While C$_3$F$_8$ is the only target fluid budgeted for PICO-500, the implications of the present CE$\nu$NS sensitivity projection study may strengthen the motivation for using a liquid argon target on a similar detector scale in the future.

All of the targets considered in this study would serve the purpose of CE$\nu$NS detection even better in case of a more closely located supernova progenitor; for example, the explosion of Betelgeuse at about \SI{222}{pc} \cite{betelgeuse} would give approximately 2,000 times higher rates in \SI{1}{t} of C$_3$F$_8$ than those presented in \autoref{tab:yield_table}. In addition, scattering rates would naturally increase with the target mass. In particular, given equal distances to a supernova progenitor, the rates in a PICO-like detector filled with \SI{25}{t} of C$_3$F$_8$ would be comparable to those in a \SI{15}{t} xenon-based detector above a \SI{5}{keV} detection threshold, potentially achievable by the XENONnT, LZ, and DARWIN collaborations in the future \cite{lang_supernova_2016,schumann_baudis_xenon_proj}. As per \autoref{tab:yield_table}, a \SI{25}{t} C$_3$F$_8$-based bubble chamber would be expected to observe up to 3 events from a supernova at \SI{50}{kpc}. The discussion of feasibility of such scaling is therefore worthwhile and will be continued within the collaboration throughout the PICO-500 program.

Taken together, these conclusions bring additional value to the C$_3$F$_8$-filled PICO-500, planned as a direct dark matter search detector. They also boost the general scientific motivation behind future design and commissioning of larger-scale detectors, as the latter would be efficient in detecting neutrinos from much more distant supernovae. 

\section{Acknowledgements}\label{sec:acknowledgements}
We thank Robert Bollig for generating the neutrino signal data, as well as Hans-Thomas Janka and Tobias Melson for making it available via Garching Core-Collapse Supernova Archive \cite{garching_archive} and participating in the related discussion; Kate Scholberg and Louis Strigari for giving helpful comments on the manuscript and suggesting further directions; Irene Tamborra and Shayne Reichard for facilitating the cross-checks of our results with those projected for xenon-filled chambers \cite{lang_supernova_2016}; and Benjamin Broerman for independent local verification of the time-integrated projections for the PICO-500 event rates. The work of TK is funded by the Undergraduate Research Initiative (URI) Stipend at the University of Alberta. We also wish to acknowledge the support of the Natural Sciences and Engineering Research Council of Canada (NSERC) and the Canada Foundation for Innovation (CFI) for funding.


\bibliographystyle{elsarticle-num.bst}


 \end{document}